\documentclass{iopart}
\usepackage{iopams}
\usepackage{graphicx}
\usepackage{units}
\usepackage{subfig}
\usepackage{ulem}
\usepackage{color}
\usepackage[latin1]{inputenc}
\usepackage{cite}
\usepackage{upgreek}
\begin{document}

\title{Indication for dominating surface absorption in crystalline silicon test masses at \unit[1550]{nm}}

\author{Alexander Khalaidovski, Jessica Steinlechner and Roman Schnabel}

\address{Institut f\"ur Gravitationsphysik,\\
Leibniz Universit\"at Hannover and Max-Planck-Institut f\"ur\\
Gravitationsphysik (Albert-Einstein-Institut), Callinstr. 38,\\
30167 Hannover, Germany}

\ead{roman.schnabel@aei.mpg.de}

\date{\today}

\begin{abstract}
The sensitivity of future gravitational wave (GW) observatories will be limited by thermal noise in a wide frequency band. 
To reduce thermal noise, the European GW observatory Einstein GW Telescope (ET) is suggested to use crystalline silicon test masses at cryogenic temperature and a laser wavelength of $\unit[1550]{nm}$.
Here, we report a measurement of the optical loss in a prototype high-resistivity crystalline silicon test mass as a function of optical intensity at room temperature. 
The total loss from both the bulk crystal \textit{and} the surfaces was determined in a joint measurement.
The characterization window ranged from small intensities below $\unit[1]{W/cm^2}$, as planned to be used in ET, up to $\unit[21]{kW/cm^2}$. 
A non-linear absorption was observed for intensities above a few $\unit[]{kW/cm^2}$. 
In addition we have observed an intensity-independent offset that possibly arises from absorption in the crystal surfaces. 
This absorption was estimated to $\alpha_{\rm{surf}}\approx\unit[800]{ppm/surface}$, which might be too high for a cryogenic operation of a fibre-suspended silicon test mass.
Such an offset was not observed in other recent measurements that were insensitive to surface absorption. 
Finally, a set of further characterization measurements is proposed to clearly separate the contributions from the surfaces and the bulk crystal.
\end{abstract}

%

\section{Introduction}
At present the construction phase of the so-called \textit{Advanced GW detectors}~\cite{ligo,virgo,kagra} nears its completion. The establishment of a gravitational wave (GW) \textit{astronomy}, however, requires a sensitivity improvement by another order of magnitude over the entire observation band.
This motivated a design study for an European 3rd-generation GW observatory, the \textit{Einstein GW Telescope} (ET)~\cite{et}, as well as first concrete considerations for upgrades of the Advanced LIGO detectors~\cite{adhikari13}. The main sensitivity limit of the advanced detector generation at mid-frequencies will be thermal noise that originates from thermally excited motions in the mirror substrates, the dielectric reflective coatings and the suspensions. To overcome these noises, it was proposed to cool the test masses of the ET low-frequency interferometer to cryogenic temperatures and to employ crystalline silicon as test mass material~\cite{et}. 
The advantage of crystalline silicon is that it is available in the dimensions required for the ET test masses.
A ``drawback" is that with an absorption coefficient of $\alpha \approx \unit[10]{/cm}$ at $\unit[1064]{nm}$ it is completely opaque at the wavelength that is currently used in GW detectors. The employment of crystalline silicon will thus enforce a change of the operation wavelength, for example to $\unit[1550]{nm}$~\cite{et} that is located within the silicon absorption gap. 

While predictions basing on photocurrent measurements led to the hope for sub-ppm/cm absorption at $\unit[1550]{nm}$~\cite{keevers95,green08}, to the best of our knowledge only recently first direct sensitive measurements of the optical absorption of undoped samples at 1550\,nm were presented~\cite{steinlechner13, degallaix12}. In~\cite{steinlechner13} we have reported an optical absorption of $\alpha = \unit[264]{ppm/cm}$. This value is two orders of magnitude higher than expected from the residual doping of the silicon crystal (the specific resistivity was specified to $\unit[11]{k\Omega cm}$). This measurement comprised both the absorption in the bulk material as well as the surface absorption. In contrast, measurements done by the LMA group in France on a $\approx\unit[30]{k\Omega cm}$ crystal that used a beam deflection method~\cite{jackson81} have found the absorption value of just the bulk crystal to be much lower in the order of $\lesssim \unit[5]{ppm/cm}$~\cite{degallaix12}. Furthermore, the measurements from the LMA group indicated that the optical absorption increases at high optical intensities.

\begin{figure}
	\centering
		\includegraphics[width=13cm]{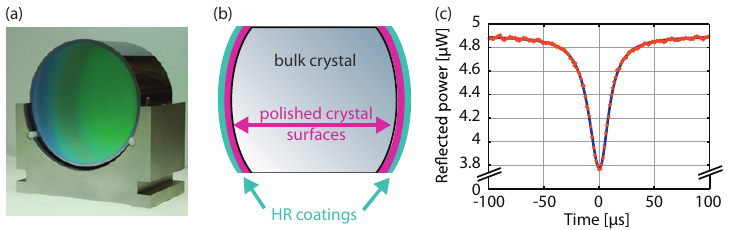}
		\caption{(a) Photograph of the prototype test mass, the diameter is 100\,mm. (b) Schematic of the monolithic cavity. Regions where optical absorption takes place, namely the bulk crystal, the polished crystal surfaces and the HR dielectric coatings, are highlighted. Independent measurements have shown that the contribution of the HR coating to the total absorption is, in a first-order approximation, negligible~\cite{coatings,steinlechner.phd}. (c) Laser power reflected by the cavity (orange dotted line) and fit of the resonance peak (solid blue line) for an intensity of $\unit[0.4]{W/cm^2}$. Without the optical loss the cavity would be impedance-matched and the minimum of the resonance peak would drop to zero, since the HR-coatings have the same reflectivity (confirmed in an independent measurement).}
	\label{fig:1}
\end{figure}

Here, we report a measurement of the optical loss in a prototype crystalline silicon test mass as a function of optical intensity. 
The measurement was performed in a silicon crystal that was highly reflectively (HR) coated at its surfaces and thus formed a monolithic cavity (Fig.~\ref{fig:1}\,(a)). 
Thus, the absorption in the bulk crystal, the dielectric coatings as well as in the surfaces beneath the coatings could be measured simultaneously as illustrated in Fig.~\ref{fig:1}\,(b). 
The characterization window ranged over more than 4 orders of magnitude from small intensities below $\unit[1]{W/cm^2}$, as planned to be used in ET, up to $\unit[21]{kW/cm^2}$. 
Our measurements confirmed the occurrence of intensity-dependent effects for intensities above a few $\unit[]{kW/cm^2}$. 

The high intensity-independent offset of $\approx\unit[250]{ppm/cm}$ observed in the measurements reported here can't be explained by the absorption in the bulk crystal. For a sample with a specific resistivity of $\unit[11]{k\Omega cm}$ this free-carrier absorption is in the order of a few ppm/cm as shown in Fig.~\ref{figure3}. This order of magnitude has also been observed by the LMA group, where a bulk absorption of $\lesssim \unit[5]{ppm/cm}$ was measured~\cite{degallaix12}. Furthermore, the contribution of the dielectric SiO$_2$/Ta$_2$O$_5$ coatings to the total absorption is with about $\unit[25]{ppm/coating}$ in a first-order consideration negligible~\cite{coatings}. This implies that the optical absorption of crystalline silicon needs to be treated separately as (a) the absorption in the bulk crystal and (b) a much higher absorption of about $\unit[800]{ppm}$ per transmission in the crystal surfaces.

\section{Method summary}
We performed absorption measurements on a silicon crystal at a wavelength of 1550\,nm. First absorption results measured with an optical intensity (here and in the following \textit{intensity} refers to the peak intensity of the Gaussian laser beam power distribution inside the cavity) of $\unit[2.2]{kW/cm^2}$ were published in~\cite{steinlechner13}. The cylindrical silicon crystal (shown in Fig.~\ref{fig:1}\,(a)) that had a diameter of 100 mm and a length of 65 mm was manufactured with the Czochralski technique. The specific resistivity was specified by the manufacturer to about $\unit[11]{k\Omega cm}$ and was due to a residual boron contamination, thus meaning an impurity concentration of approximately $\unit[2\times 10^{12}]{atoms/cm^3}$~\cite{bor}. The crystal end surfaces were polished to be convex curved with a radius of curvature (ROC) of $\unit[1]{m}$ and coated with a highly reflective SiO$_2$/Ta$_2$O$_5$ coating using ion beam sputtering. 
The power reflectivity at the wavelength of 1550 nm was measured to R$_1$\,=\,R$_2$\,=\,$99.96$.
Thus, a monolithic optical cavity with a free spectral range (FSR) of 663 MHz and a finesse of $\approx 1500$ was available for absorption measurements.

In~\cite{steinlechner13} we have shown that the optical absorption per round-trip and the total cavity round-trip loss (L$_{\rm{RT}}$) were identical within a few percent, which corresponded to the measurement error bars. Thus, a simple measurement of the light's impedance matching to the cavity (with known mirror reflectivities) provides the optical absorption per round trip.
For this measurement, a 1550 nm laser beam was matched to the eigenmode of the monolithic cavity and detected in reflection. The laser frequency was linearly varied in the vicinity of a cavity resonance.
Frequency markers imprinted with an electro-optic modulator provided an accurate calibration of the time axis.
A more detailed description of the measurement setup is given in~\cite{steinlechner13}.

The laser power injected into the cavity was varied from $\unit[5]{\upmu W}$ to $\unit[700]{mW}$, being equivalent to intra-cavity peak intensities of $\unit[0.4]{W/cm^2}$\,-\,$\unit[21]{kW/cm^2}$, respectively. For each input power we performed 10 measurements of the cavity resonance peak using different scanning frequencies to ensure that the measurement was not limited by time-dependent effects. The peaks were fitted using the software developed for the photothermal self-modulation technique~\cite{SHG} and the impedance mismatch was determined (please note that the photothermal self-modulation technique itself was not applied for the measurements reported in this paper since at optical powers $\lesssim \unit[1]{kW/cm^2}$ no absorption-induced deformation of the resonance peaks that is required for the method took place).
The fitting parameters were the cavity in-coupling power reflectivity R$_1$ and the  effective out-coupling reflectivity $\widetilde{R_2}= R_2 + \rm{L}_{\rm{RT}}$. $\widetilde{R_2}$ accounts for the entire cavity round-trip loss except for the transmission of the in-coupling coating. 
From this fit the round-trip loss was derived, the results are shown in Fig.~\ref{figure2} and discussed in the following section. Further geometric and material parameters used in the simulation were the crystal length of $\unit[65]{mm}$ and the silicon index of refraction $\textrm{n}_{\rm{Si}}=3.48$~\cite{frey06}. Figure~\ref{fig:1}\,(c) shows an exemplary measurement at an intensity of $\unit[0.4]{W/cm^2}$ and a fit of the resonance peak (orange dotted line and solid blue line, respectively).

\section{Discussion}
\begin{figure}[t]
	\centering
		\includegraphics[width=12cm]{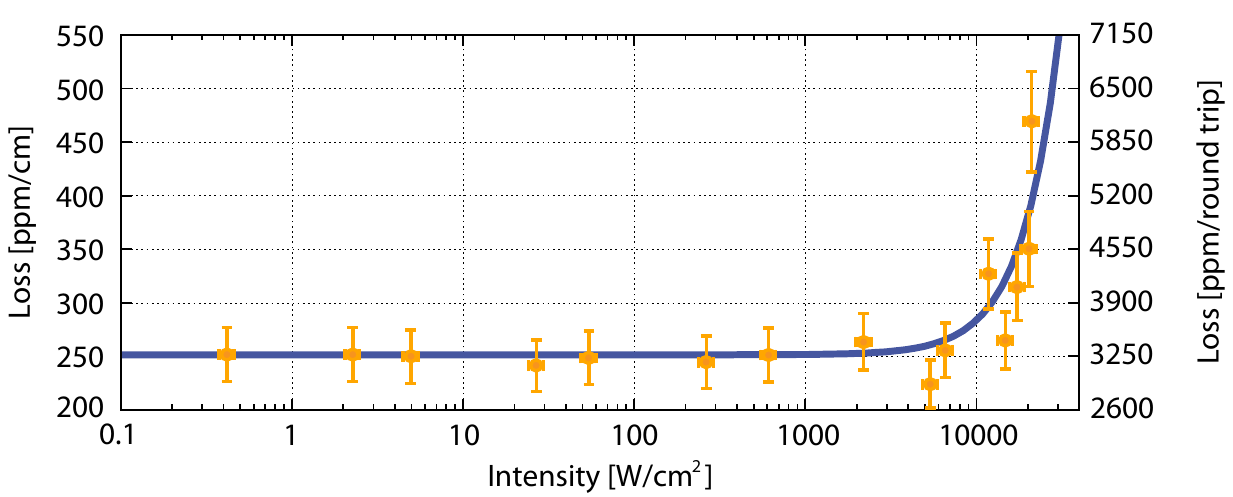}
		\caption{Optical absorption in our test mass derived from the impedance mismatch measurement as a function of the intra-cavity peak intensity. Left y-axis: absorption values scaled to the round-trip length. Right y-axis: total cavity round-trip loss.}
	\label{figure2}
\end{figure}
\begin{figure}[b]
	\centering
		\includegraphics[width=12cm]{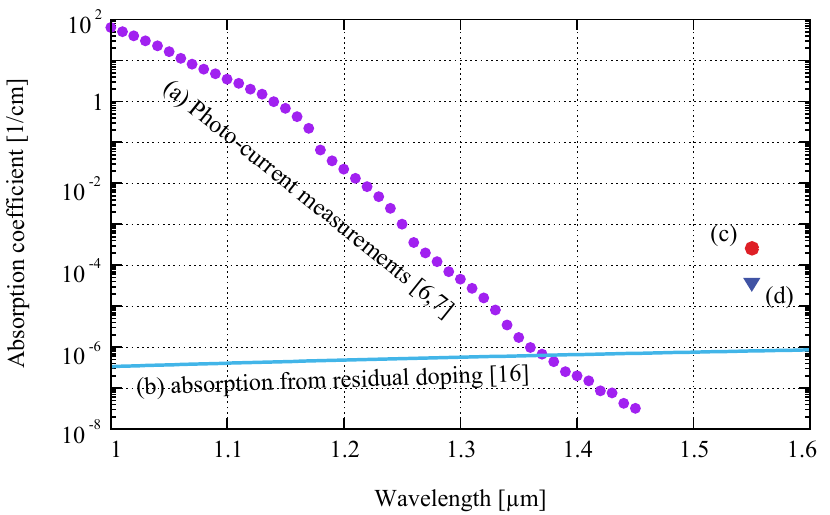}
		\caption{Comparison of absorption values of crystalline silicon in the near infrared band. 
		(a) Band-band absorption derived from photocurrent measurements~\cite{keevers95,green08}. 
		(b) Free-carrier absorption calculated for the residual doping in our sample (p-doping, $N=\unit[2\times 10^{12}]{/cm^3}$) using the theory presented in~\cite{soref86}. 
		(c, red circle) Total optical absorption in a prototype crystalline silicon test mass measured at an intensity of $\unit[2.2]{kW/cm^2}$ and scaled to the cavity round-trip length~\cite{steinlechner13}. 
		(d, head-down blue triangle) Bulk absorption of a silicon crystal measured with a beam deflection approach at an intensity of $\unit[2.2]{kW/cm^2}$~\cite{degallaix12}. 
}
	\label{figure3}
\end{figure}

Figure~\ref{figure2} shows the results obtained for the optical absorption and a least-squares fit ($y=3.2\cdot10^{-7}x^2+2.7\cdot10^{-9} x+251.4$) of the measurement distribution (orange dots and solid blue line, respectively). The left y-axis provides absorption values scaled to the round-trip length of 13\,cm (thus in ppm/cm), while the right y-axis gives the total round-trip loss. Following the discussion given in~\cite{steinlechner13}, a maximal error of 10\%, resulting from uncertainties in the simulation input parameters, the cavity geometric parameters and the mode-matching was assumed. As a consistency check, the measurements reported here were compared to the absorption value obtained for an intensity of $\unit[2.2]{kW/cm^2}$ using the photo-thermal self-phase modulation technique~\cite{steinlechner13} (shown in trace (c) of Fig.~\ref{figure3}). Both results excellently agree within the error bars, giving each time a value of $\alpha=\unit[264]{ppm/cm}$.

The results of the measurement shown in Fig.~\ref{figure2} are twofold. First, our measurements show the impedance matching of the cavity to change in a non-linear way as a function of optical intensity. 
Assuming that this effect is due to bulk absorption, our results thus qualitatively confirm the measurements reported in~\cite{degallaix12} by the LMA group that used a beam deflection approach~\cite{jackson81}, although the non-linear increase of the absorption was observed at an intensity lower by more than one order of magnitude in the LMA measurements.
Degallaix \textit{et al.}~explained the non-linear behaviour by two-photon absorption processes.
Such two-photon absorption creates free carriers that strongly increase the loss due to the free-carrier absorption when compared to just the absorption due to the residual dopant concentration. 
The free-carrier density and thus the effective absorption, however, depends on the free carrier lifetime and hence on the silicon crystal purity. 
As Degallaix \textit{et al.}~have shown in~\cite{degallaix12}, samples with higher purity (and thus a longer free-carrier lifetime) demonstrated a higher absorption for the same intensity.
While the LMA measurements used a float-zone purified crystal with a specific resistivity of $\approx\unit[30]{k\Omega cm}$, our crystal was Czochralski-grown with a specific resistivity of about $\unit[11]{k\Omega cm}$. Although a detailed analysis of the free-carrier absorption as a function of specific resistivity is required, it seems likely that the disagreement in power dependence is due to the difference in the free carrier lifetime. 

Second, the absorption results of $\gtrsim \unit[250]{ppm/cm}$ reported here and in~\cite{steinlechner13} are significantly higher than the results from other groups as shown in Fig.~\ref{figure3}\,(a,d). The measurement approaches employed there consisted in (i) spectral response measurements on solar cells~\cite{keevers95} and (ii) beam deflection measurements~\cite{degallaix12}. The first case only provides an indirect measurement, its results are therefore not unconditionally transferable to optical absorption and may be rather used to derive a lower limit as discussed in~\cite{steinlechner13}.
While the beam deflection approach is generally much better suited to characterize the optical absorption, it only provides information about the volume where the probe beam interacts with the thermal lens created by a strong pump beam. 
In~\cite{degallaix12} this region was selected to be within the optical substrate. Thus, merely the bulk absorption was evaluated, while no full information about the optical absorption of a complete test mass including both the bulk crystal \textit{as well as the surfaces} was obtained.

In contrast, our measurement approach optimally reflects the real application in a GW observatory, because it provides the full information about the optical round-trip loss. 
Both the bulk absorption and the absorption in the surfaces beneath the HR coatings contribute to this loss as illustrated in Fig.\ref{fig:1}\,(b). When compared to the LMA measurements, it stands out that at low intensities our data converges to an absorption coefficient of $\alpha=\unit[250]{ppm/cm}$ while the LMA data indicates $\alpha$ to be in the order of $\lesssim \unit[5]{ppm/cm}$. Apart from that the two curves show a very similar characteristic. At low intensities no non-linear contribution to the free-carrier absorption takes place. As outlined in~\cite{steinlechner13}, from the residual doping of the crystal an absorption coefficient of $\alpha\approx\unit[1]{ppm/cm}$ can be predicted, excluding this effect as a possible explanation for the observed high absorption values. The expected contribution is shown in trace (b) of Fig.~\ref{figure3} (solid line).

A possible explanation for the difference between our observations and the LMA data~\cite{degallaix12} is absorption in the surfaces of the prototype test mass. The crystal surface was super-polished to a specified residual microroughness of  $\unit[2-4]{\mathring{A}}$. It is known that the polishing procedure can generate thin layers of amorphous silicon (a-Si) on the crystal surface~\cite{daum}. At the wavelength of $\unit[1550]{nm}$ the absorption of amorphous silicon is many orders of magnitude larger than the absorption of crystalline silicon. Although the absorption coefficient depends on the thermal treatment of the material, so that in contrast to the crystalline material no unique value can be given, absorption values of $\unit[100]{/cm}$ up to $\unit[2000]{/cm}$ at 1550\,nm (0.8\,eV) have been reported~\cite{brodsky,loveland,chittick}. It is worth noting that very few information is available on pure amorphous silicon, since the semiconductor industry merely concentrates on hydroginated material (a-Si:H) where a vast range of literature is available (see e.\ g.~\cite{aSilicon}). 
In a different experiment, we measured the absorption in a HR stack of Si/SiO$_2$ dielectric coatings to a value of 1000\,ppm - 1500\,ppm, depending on the polarization of the probe beam. The Si-layers of this coating also (mainly) consist of amorphous material. 
Since in a HR coating the penetration depth of the laser beam typically only comprises the first few layers, the absorption can be estimated to a value of several tens/cm. This value is even smaller than expected for amorphous silicon. However, since thermal treatment and the procedure used to apply the Si film strongly influence the absorption value (indeed a recrystallization can be achieved by tempering) no reliable expectation value can be derived from literature. A publication is currently in preparation~\cite{steinlechner13b}.
For the sake of completeness it shall be mentioned that after polishing a layer of native oxide forms on the surface and reaches a thickness of 1-2\,nm after several weeks-months, strongly depending on the environmental conditions~\cite{native-oxide}. This layer, however, mainly consists of SiO$_2$ and thus in a first order estimate is not expected to significantly contribute to the absorption.

\begin{figure}[t]
	\centering
		\includegraphics[width=12cm]{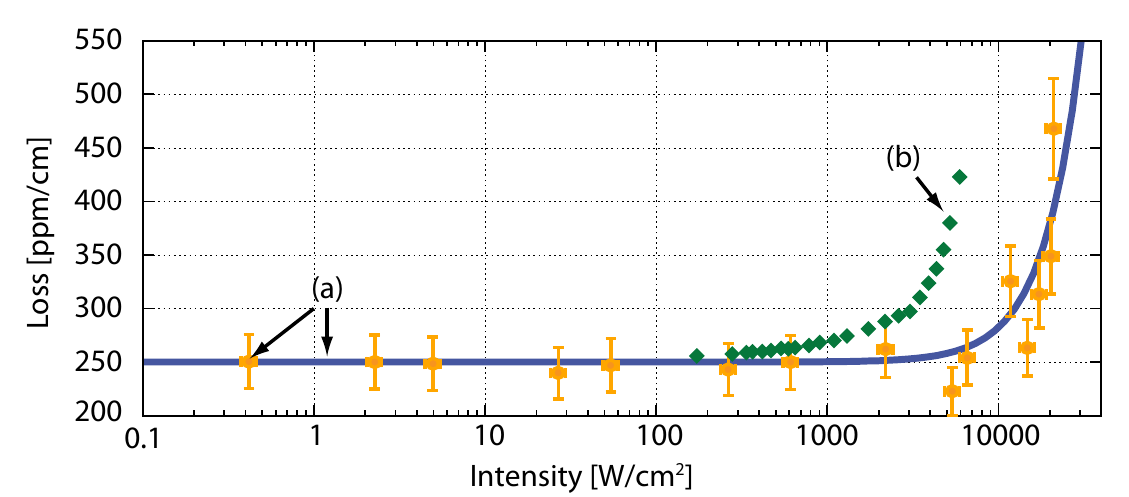}
		\caption{(a) The data shown in Fig.~\ref{figure2} replotted for a comparison to trace (b) taken from~\cite{degallaix12} that shows the bulk absorption obtained from beam deflection measurements on a silicon crystal with a specific resistivity of $\approx\unit[30]{k\Omega cm}$. According to the discussion given in the text, an offset of $\alpha_{\rm{off}}=\unit[250]{ppm/cm}$ was added to the data from ref.~\cite{degallaix12}.}
	\label{figure4}
\end{figure}

In view of the above argumentation, for low intensities as planned to be employed in ET our measurement results can be interpreted as follows: The total absorption consists of
\begin{enumerate}
\item a bulk absorption $\alpha_{\rm{bulk}}$ of a few ppm/cm that is negligible when compared to
\item an offset of about $\alpha_{\rm{off}}\approx\unit[250]{ppm/cm}$ due to absorption in the crystal surfaces. Further measurements for a more exact determination are proposed in section~\ref{sec:conclusion}.
\end{enumerate}
Considering the fact that during one cavity round-trip four surface transmissions take place (see Fig.\ref{fig:1}\,(b)), the absorption per surface can thus be estimated to $\unit[250]{ppm/cm}\times\unit[13]{cm}/4\approx\unit[800]{ppm}$. 
Except for this offset our data shows a very good agreement with the LMA measurements. For comparison, Fig.~\ref{figure4} shows our results along with the LMA data to which a constant offset of  $\alpha_{\rm{off}}=\unit[250]{ppm/cm}$ was added.
The very similar characteristics support the assumption of a constant offset that cannot be explained by impurities resulting from residual doping. 
It strikes out that in the LMA measurements the increase of the absorption sets in at lower intensities than in our case. As discussed above, a possible explanation for this is the higher specific resistivity of the material analyzed by the LMA group that determines the lifetime of the generated free carriers.

\section{Conclusions and outlook}
\label{sec:conclusion}

Summarizing, our measurements confirmed the non-linear dependence of the absorption in crystalline silicon on the optical intensity reported by Degallaix \textit{et al.}~\cite{degallaix12}. In addition to the bulk crystal absorption we have observed an intensity-independent offset that possibly arises from absorption in a-Si surfaces created during the polishing procedure. In a separate measurement~\cite{coatings} it was shown that the absorption in the dielectric SiO$_2$/Ta$_2$O$_5$ coatings is orders of magnitude smaller and therefore negligible for this consideration.
Our results imply an absorption loss of  $\approx\unit[1600]{ppm}$ for a single transmission through a complete test mass, e.\,g.\ through the input test masses of the ET arm cavities.

While ET will use large beam diameters and thus rather low optical intensities~\cite{et} so that the non-linear contributions to the absorption will be negligible, the surface absorption might become an issue. The laser power absorbed in the test masses will have to be extracted from the system via the suspensions. Because of the small fiber diameters only a very limited amount of thermal energy can be extracted. Since the total amount of generated heat determines the lower limit for the test mass temperature that can be achieved, the high surface absorption will possibly restrict the suppression of thermal noise in ET. Therefore, a series of further characterization measurements is currently in preparation. In a first step, a set of monolithic cavities with different lengths will be prepared and characterized. These results will allow to precisely separate the bulk absorption from the surface absorption, since the latter will only lead to a constant offset independent from the cavity length. Thereupon crystals with a different specific resistivity ranging from $\unit[1.5]{k\Omega cm}$ to $\unit[70]{k\Omega cm}$ will be compared to analyze the influence of the residual doping on the optical absorption. The goal of these measurements will be to analyze whether crystals produced with the Czochralski technique are usable in ET in terms of the optical absorption. Today Czochralski-grown crystals that fulfill the ET requirements (450\,mm diameter) can already be produced. In contrast it is not yet clear whether the considerably purer float zone grown crystals will be available in the same dimensions within the next decade due to technical difficulties intrinsic to the manufacturing process. Thus, a confirmation of the fact that Czochralski-grown crystals fulfill the ET absorption requirements would significantly relax the requirements on the ET test masses in terms of availability and feasibility. Finally, the discussed measurements will be repeated at the ET design temperature of 10\,K to characterize the influence of cryogenic temperatures on the free-carrier absorption.

\section*{Acknowledgements}

The autors would like to thank W. Daum, J. Degallaix, H. Lück and T. Wietler for intense and helpful discussions. 
We acknowledge support from the SFB/Transregio 7, the International Max Planck Research School (IMPRS) on Gravitational Wave Astronomy, and from QUEST, the centre for Quantum Engineering and Space-Time Research.

\end{document}